\documentclass[twocolumn, prb]{revtex4-1}
\usepackage{amssymb} 
\usepackage{amsmath}
\usepackage{epsfig}
\usepackage{graphics, graphicx}
\usepackage{bbold}
\usepackage{psfrag}
\usepackage{mathcomp}
\usepackage{subfigure}
\usepackage{verbatim}
\usepackage{color}
\usepackage[colorlinks,citecolor=blue]{hyperref}

\begin{document}

\title{Magnetic Manifestation of Discrete Scaling Symmetry in Dirac Semimetals}
\author{Mingyuan Sun}
\email{msun@connect.ust.hk}
\affiliation{Institute for Advanced Study, Tsinghua University, Beijing, 100084, China}

\date{\today}

\begin{abstract}
Two-dimensional and three-dimensional massless Dirac fermions can form a sequence of quasibound states with an attractive charged impurity. These quasibound states exhibit a discrete scaling symmetry, i.e., the energy ratio between two successive states is a constant. Through the calculation of the energy spectrum directly, we find that in two dimension an applied magnetic field can shift or even destroy the quasibound states around the Dirac point and their discrete scaling symmetry disappears. However, as the magnetic field increases, the remaining quasibound states are pushed up to the Dirac point. When one quasibound state is close to the Dirac point, the spectrum is modified significantly, due to the resonant scattering. The magnetic oscillation of the spectrum displays the same discrete scaling symmetry as the quasibound state does in zero magnetic field. This phenomenon also occurs in the quantum limit of three-dimensional Dirac semimetals, where the system becomes quasi one-dimensional essentially. Our theoretical analysis are in good agreement with the recent experimental observations. 
\end{abstract}
\maketitle

\section{Introduction}

Continuous scaling symmetry can be broken into a discrete scaling symmetry (DSS), with an appropriate boundary condition. One famous example is the Efimov effect, which was first discovered by Vitaly Efimov\cite{Efimov}.  Three identical bosons can form a sequence of bound states in the vicinity of two-body s-wave resonance. The energies of the bound states obey a discrete geometric scaling law, i.e., exhibiting a DSS. Here, the continuous scaling symmetry of the corresponding Schrodinger equation is broken by a short-range boundary condition. The Efimov effect has been observed in cold atoms\cite{Efimov_Exp0,Efimov_Exp1,Efimov_Exp2,Efimov_Exp3,Efimov_Exp4,
Efimov_Exp5,Efimov_Exp6,Efimov_Exp7} as well as in helium\cite{Efimov_Exp8}. Another interesting example is the Efimovian expansion of the unitary fermi gas, where a DSS appears in the dynamics of the many-body system\cite{Efimov_Dyn1, Efimov_Dyn2, Efimov_Dyn3}. 

Dirac and Weyl semimetals have been extensively studied in condensed matter physics\cite{Dirac1, Dirac2, Dirac3}. As an example, atomic collapse in a Coulomb potential has been investigated both theoretically and experimentally in graphene, due to its large ``fine structure constant" and zero gap\cite{AC1,AC2,AC3,AC4,AC5,AC6,AC7,AC8,AC9,AC10,AC11,AC12}. The massless Dirac equation with an attractive Coulomb potential possesses a continuous scaling symmetry. When a short-range boundary condition is added, two or three dimensional (2D or 3D) massless Dirac fermions can form quasibound states, which also display a DSS\cite{AC4,AC6,AC12,DiracEfimov1,DiracEfimov2,DiracEfimov3,DiracEfimov4}. However, it is until recently that some signatures of them have been observed experimentally\cite{Wang, Akkermans}. Ovdat \textit{et al.} measured the differential tunnel conductance around a charged impurity in graphene and extracted the spectrum directly\cite{Akkermans}. Wang \textit{et al.} observed a log-periodic magnetoresistance oscillation in the ultra-quantum limit of a potential 3D Dirac semimetal ZrTe$_5$\cite{Wang}. When a magnetic field is applied, a new length scale is implemented to the system and it could destroy the DSS of the original quasibound states. Thus a complete theory is needed to describe the interplay between Efimov-like quasibound states and the magnetic field in these systems.

\begin{figure}[b]
\includegraphics[width=7.0cm,height=5.9cm]{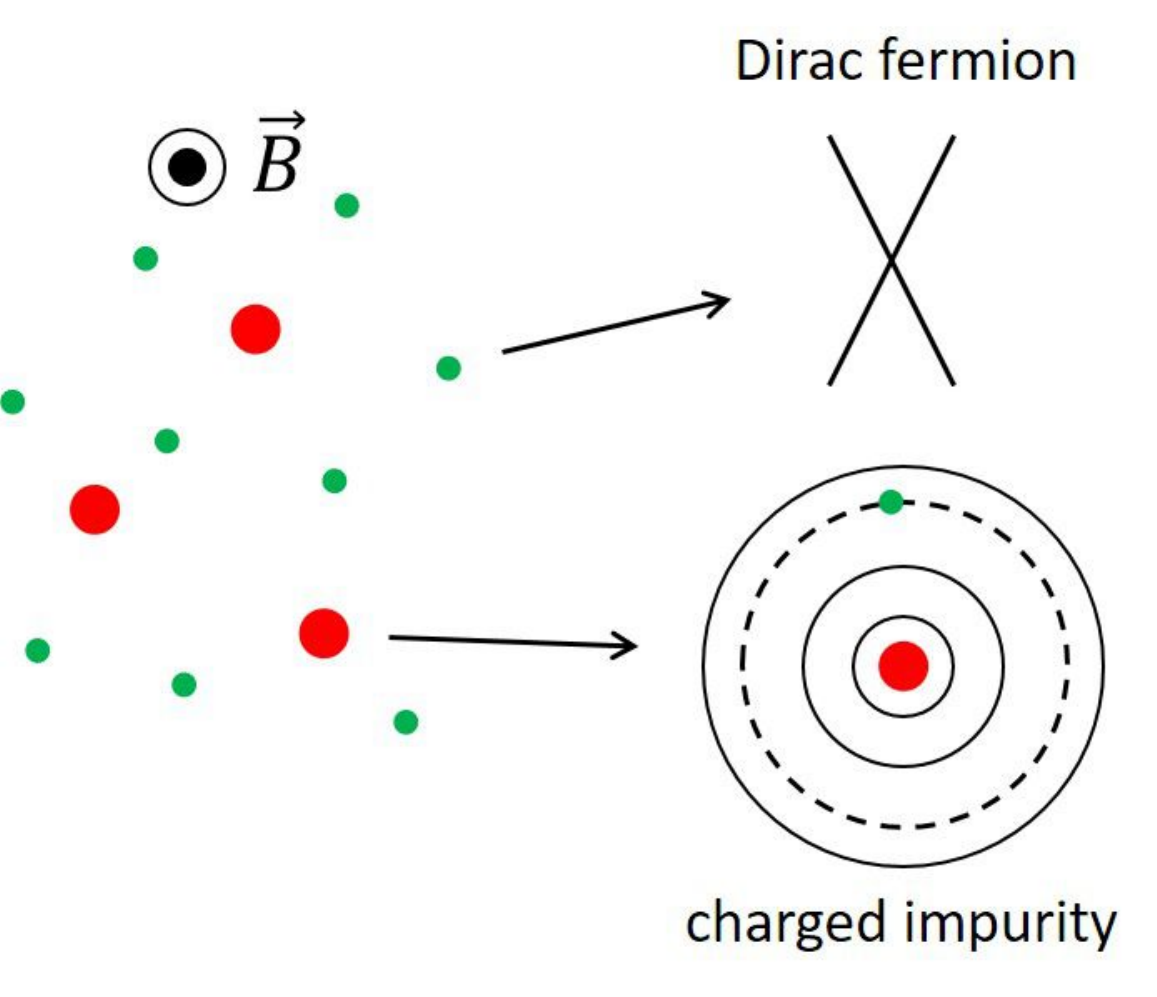}
\caption{(Color Online). Schematics of massless Dirac fermions (green) with attractive charged impurities (red) in a magnetic field. Massless Dirac fermions can form a sequence of quasibound states (solid black circles) with attractive charged impurities, which display a discrete scaling symmetry. In a magnetic field, Landau levels (dashed black circle) appear. In this paper, we take the symmetric gauge and the low-density limit of impurities, as well as the location of the impurity being the original point. The interplay between Efimov-like quasibound states and Landau levels could lead to new phenomena.
\label{schematics}}
\end{figure}

In this paper, we study 2D and 3D massless Dirac fermions interacting with attractive charged impurities in a magnetic field, as shown in Fig.~\ref{schematics}. By calculating the energy spectrum directly, we show that, in 2D the DSS of the Efimov-like quasibound states is destroyed around the Dirac point, by the applied magnetic field. However, as the magnetic field increases, the implemented new length scale (i.e., the magnetic length) can move the deep quasibound states up to the Dirac point. When one is close to the Dirac point, the spectrum varies drastically, due to the resonant scattering between Landau level's states and that quasibound state.  Thus, the spectrum oscillates and the period recovers the original DSS. A similar phenomenon can also occur in the quantum limit of 3D Dirac semimetals, where the strong magnetic field constrains all the carriers at the lowest Landau level and tranforms the system into quasi-1D.  Our results agree well with the experimental observations\cite{Wang}.

\section{Discrete scaling symmetry in 2D Dirac semimetals}

Since we study the effect of quasibound states, which are localized around the charged impurity, the interaction of other impurities can be negligible, as long as the size of the quasibound state is smaller than the distance between neighbor impurities. On the other hand, the applied magnetic field introduces another length scale, i.e., magnetic length, which can also be adjusted to be small, so as to ignore the interference between impurities. Thus, our study will focus on the low-density limit of impurities. It is equivalent to solving the corresponding Dirac equation with one charged impurity. The massless Dirac equation for electrons in a Coulomb potential and a magnetic field can be expressed as
\begin{widetext}
\begin{equation}
 \begin{pmatrix}
   0 & v_F \vec{\sigma}\cdot (\mathbf{P}+e\mathbf{A})  \\
   v_F \vec{\sigma}\cdot (\mathbf{P}+e\mathbf{A}) & 0 
 \end{pmatrix}
 \begin{pmatrix}
   \psi_1(\mathbf{r})  \\
   \psi_2(\mathbf{r}) 
 \end{pmatrix}  = (E-V(r))
  \begin{pmatrix}
   \psi_1(\mathbf{r})  \\
   \psi_2(\mathbf{r}) 
 \end{pmatrix}
  \label{DiracEq}
\end{equation}
\end{widetext}
Here, $v_F$ is the Fermi velocity. $\vec{\sigma}$ are Pauli matrices and $\mathbf{P}$ is the momentum operator. $-e$ is the electric charge of the electron. $\mathbf{A}$ and $V(r)$ are respectively the magnetic vector potential and the Coulomb potential. We choose the symmetric gauge $\mathbf{A}=\frac{B}{2}(-y,x,0)$. $V(r)=-Ze^2/(4\pi \varepsilon_0 r)$, with $Ze$ being the electric charge of the impurity and $\varepsilon_0$ being the vacuum permittivity. $\psi_{1(2)}(\mathbf{r})$ is the eigen-wavefunction with the eigen-value $E$.

In zero magnetic field, the wavefunctions $\psi_{1(2)}$ can be written in the form of partial-wave expansion. In 2D, the $4\times 4$ Dirac matrix can be decoupled into $2\times 2$ matrix. If we assume $\psi_{1(2)}(\mathbf{r})=(u_1(r)e^{i m\varphi},u_2(r)e^{i (m+1)\varphi})^T$, the radial equation can be expressed as
\begin{equation}
\frac{d}{dr}
  \begin{pmatrix}
   u_1(r)  \\
   u_2(r) 
 \end{pmatrix}=
 \begin{pmatrix}
   \frac{m}{r} & i(\frac{E}{\hbar v_F}+\frac{Z\alpha}{r}) \\
   i(\frac{E}{\hbar v_F}+\frac{Z\alpha}{r})  &  -\frac{m+1}{r}
 \end{pmatrix}
 \begin{pmatrix}
   u_1(r)  \\
   u_2(r) 
 \end{pmatrix}
 \label{Dirac2D}
\end{equation}
where $\alpha=e^2/(4\pi\epsilon_0 \hbar v_F)$ is the fine structure constant and $m$ is an integer. For $|Z\alpha| > |m+\frac{1}{2}|$, it was shown that an infinite family of Efimov-like quasibound states exist below the Dirac point in each channel $m$, with a scaling factor $e^{\pi/\sqrt{(Z\alpha)^2-(m+1/2)^2}}$\cite{AC4,AC6,AC12,DiracEfimov1,DiracEfimov2,DiracEfimov3}. 

\begin{figure}[t] 
    \includegraphics[width=7.50cm,height=9.0cm]{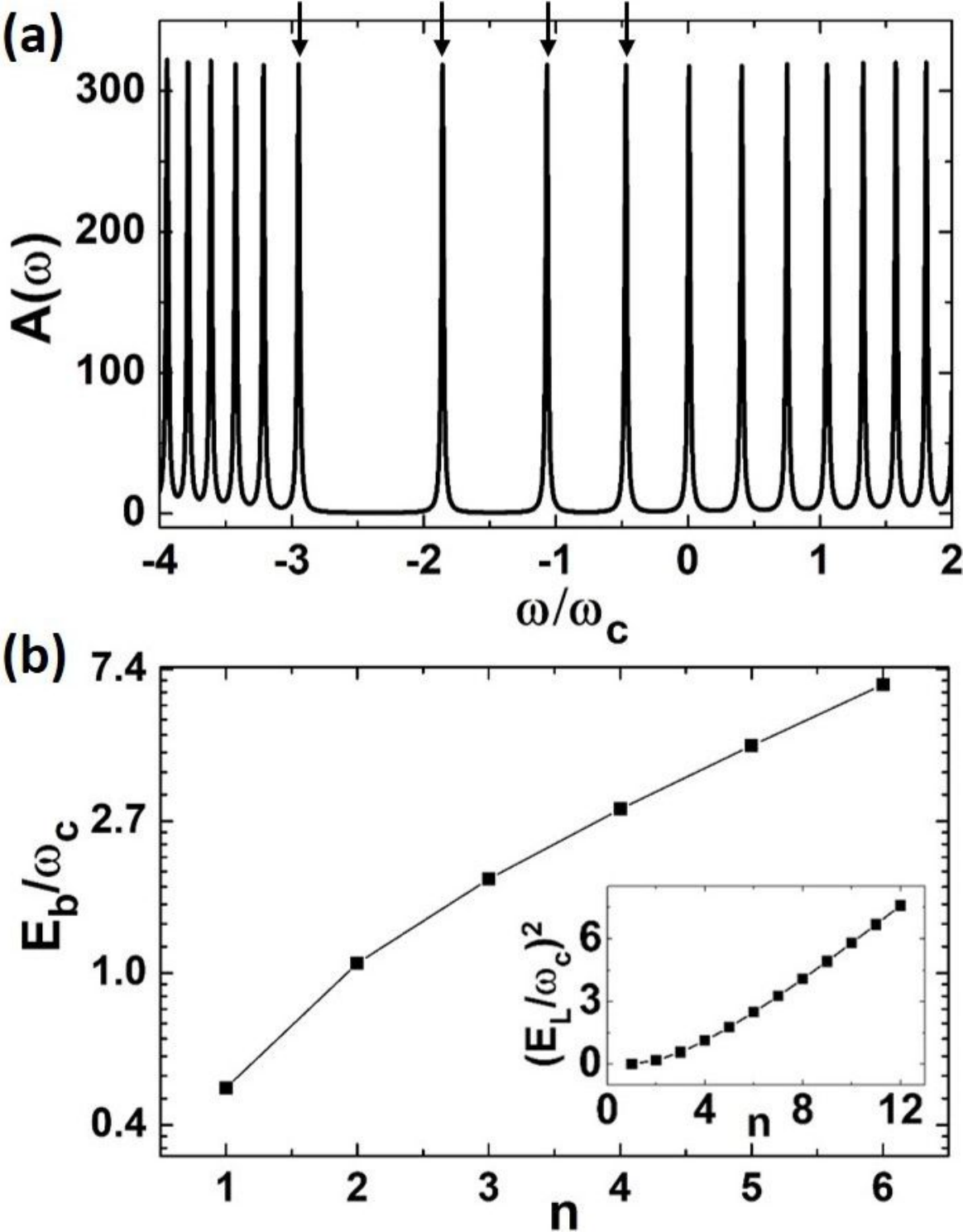}
    \caption{(Color Online). The spectrum of 2D massless Dirac electrons in an attractive Coulomb potential and magnetic field. Here, $A(\omega)$ represents the density of states at a single channel, which should be Dirac $\delta$-functions, since it is a single-particle problem. The energy positions are just the eigenvalues of Eq.~\ref{DiracEq}. A small width is added artificially to make them visible. $Z\alpha=10$ is set. $\omega_c=\sqrt{2}\hbar v_F/l_B$ is the cyclotron frequency of Dirac fermions. (a) The spectrum of the channel $m=0$. (b) The sequence of binding energies (normalized by the cyclotron frequency $\omega_c$) in a logarithmic scale. Here, the quasibound states are distinguished by the corresponding wavefunctions\cite{Sun}. Several of them around the Dirac point are labeled by vertical arrows in (a). Owing to the magnetic field, shallow quasibound states do not obey the geometric scaling law, while deep ones appear Efimov-like roughly.  Inset: the square of the normalized Landau level's energy ($\omega>0$). Lower Landau levels are significantly changed by the Coulomb potential, while higher ones still satisfy $E_n \propto \sqrt{n}$ as the free Dirac fermions. }
     \label{results2Dsf}
\end{figure}

When the magnetic field is applied without a Coulomb potential, the system will form Landau levels with wavefunctions $(\psi_{n,m} (\mathbf{r}))/\sqrt{2}, \pm i\psi_{n-1,m+1} (\mathbf{r})/\sqrt{2})^T$ for $E_n=\pm \sqrt{n}\hbar \omega_c$, ($n>0$ and $m\geq -n$. For $n=0$ Landau level, it is  $(\psi_{0,m} (\mathbf{r}),0)^T$).
\begin{align}
 \psi_{n,m}(\mathbf{r})= &\frac{1}{\sqrt{2\pi l_B^2}}\sqrt{\frac{n!}{(n+m)!}}e^{im\varphi - r^2/(4 l_B^2)} \nonumber \\
  &\cdot(\frac{r^2}{2l_B^2})^{m/2} L_n^{(m)}(\frac{r^2}{2 l_B^2})
 \label{Landauwf}
\end{align}
where, $l_B=\sqrt{\hbar/(eB)}$ is the magnetic length and $L_n^{(m)}$ is the associated Laguerre polynomial. $\omega_c=\sqrt{2}\hbar v_F/l_B$ is the cyclotron frequency of Dirac fermions. In both the magnetic field and the Coulomb potential, $m$ is still a good quantum number and different channels (labelled by $m$) remain decoupled. We use Landau levels' wavefunctions Eq.~\ref{Landauwf} as the basis, to solve Eq.~\ref{DiracEq} and obtain the spectrum, as shown in Fig.~\ref{results2Dsf}. Here, a short-range cutoff $r_c$ is employed and $V(r)=0$ for $r<r_c$. Around the Dirac point, the quasibound states do not obey the geometric scaling law, denoting the breaking DSS, due to the magnetic field (see Fig.~\ref{results2Dsf}(b)). The shallow ones are even destroyed. Similarly, Landau levels do not satisfy the relation $E_n \propto \sqrt{n}$ as the free 2D Dirac fermions\cite{Dirac1}, due to the Coulomb potential. Thus, the spectrum around the Dirac point can be significantly modified by the interplay between the magnetic field and the Coulomb potential. As a comparison, far away from the Dirac point, deep quasibound states are still roughly Efimov-like, while Landau levels obey $E_n \propto \sqrt{n}$ approximately.

\begin{figure}[t] 
    \includegraphics[width=7.50cm,height=9.0cm]{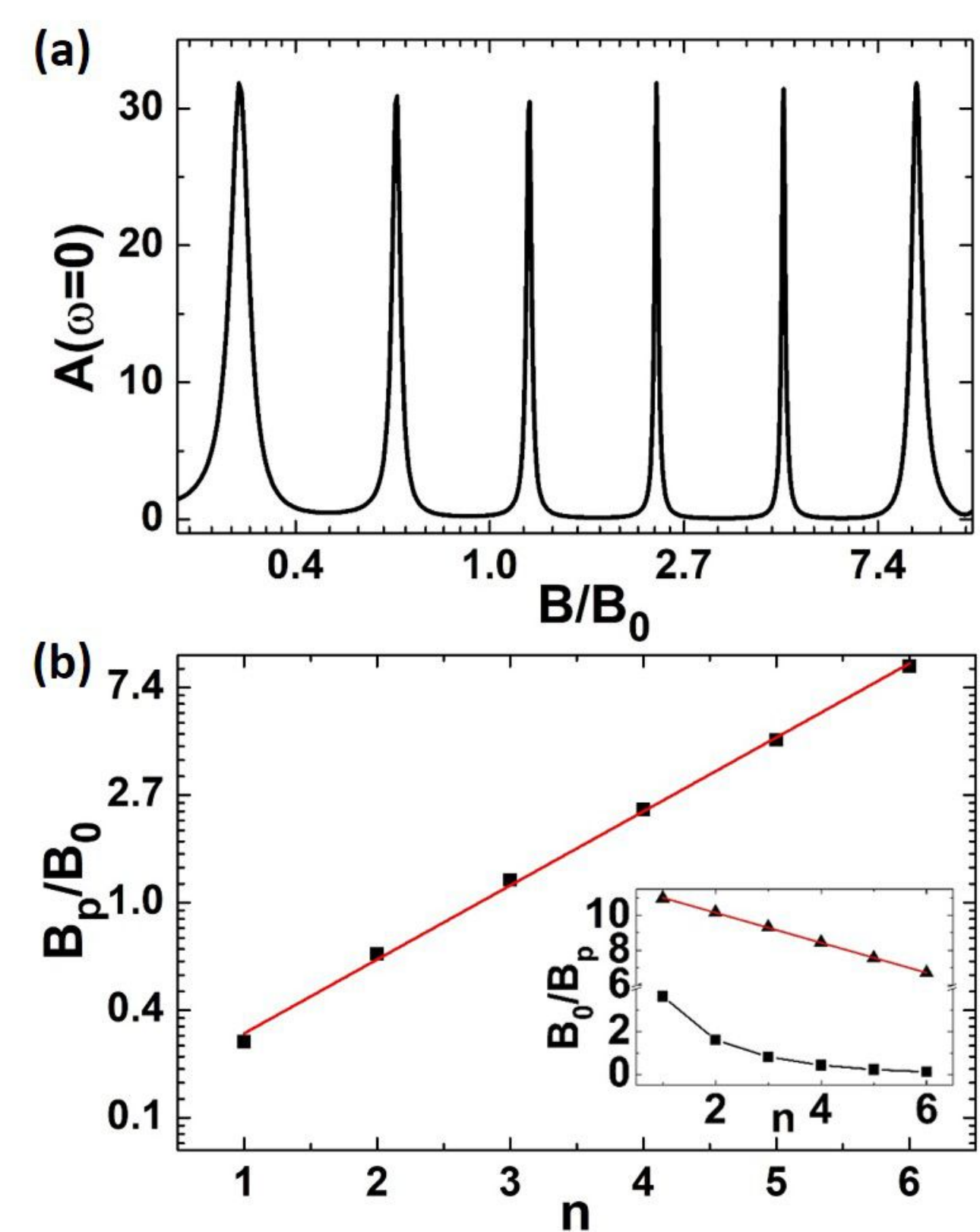}
    \caption{(Color Online). The spectrum (the density of states) for the channel $m=0$ at various magnetic fields. $Z\alpha=10$ and $B_0$ is a chosen magnetic field as a reference, with $l_{B0}/r_c=1$. Note that only the locations of the peaks are physical, since the widths are added artifically as in Fig.~\ref{results2Dsf}. (a) Magnetic dependence of the density of states $A(\omega)$ at the Dirac point in a logarithmic scale. (b) The sequence of magnetic locations (i.e., the value of the magnetic field) of the spectrum's peaks (shown in (a)) in a logarithmic scale. Red line denotes a linear fitting, which displays a geometric scaling law. As a comparison, the inverse of magnetic locations (black squares) are drawn in the inset. It clearly deviates from the $1/B$-law, which is expected for free Dirac fermions in the Shubnikov-de Haas (SdH) oscillations\cite{Dirac1}. In the inset, we also show the result (black triangles) for the energy $\omega/\omega_{c0}=1.0$ ($\omega_{c0}=\sqrt{2}\hbar v_F/l_{B0}$), which still obeys the $1/B$-law approximately (red line denoting a linear fitting). }
     \label{results2Dsfb}
\end{figure}

As the magnetic field varies, the spectrum (i.e., the density of states) at the Dirac point oscillates and the periods exhibit a DSS (see Fig.~\ref{results2Dsfb}). In a semi-classical way, the magnetic length roughly represents the size of the Landau level's orbits (i.e., wavefunctions). As the magnetic length decrease, the Landau level's orbits can scatter with deeper quasibound states of the impurity. In the energy view, it means that the quasibound states are shifted up to the Dirac point, since the lowest Landau level ($n=0$) is fixed around the Dirac point. When the quasibound state is close to the Dirac point, the electron at the lowest Landau level can scatter resonantly with the impurity, which changes the spectrum drastically. Therefore, the DSS of the original quasibound states is manifest in the magnetic oscillation of the spectrum. 

This phenomenon can be simply interpreted in the following way. When a magnetic field is applied, it introduces a new length scale, denoted by the magnetic length $l_B$. Eq.~\ref{Dirac2D} becomes
\begin{equation}
\frac{d}{dr}
  \begin{pmatrix}
   u_1(r)  \\
   u_2(r) 
 \end{pmatrix}=
 \begin{pmatrix}
   \frac{m}{r}+\frac{r}{2l_B^2} & i(\frac{E}{\hbar v_F}+\frac{Z\alpha}{r}) \\
   i(\frac{E}{\hbar v_F}+\frac{Z\alpha}{r})  &  -\frac{m+1}{r}-\frac{r}{2l_B^2}
 \end{pmatrix}
 \begin{pmatrix}
   u_1(r)  \\
   u_2(r) 
 \end{pmatrix}
 \label{Dirac2Db}
\end{equation}
The additional terms $\pm r/(2l_B^2)$ break the continuous (or discrete) scaling symmetry of the original Dirac equation (i.e., Eq.~\ref{Dirac2D}). Hence they also destroy the DSS of the original quasibound states. However, if we transform the length $r\rightarrow \eta r $ and $l_B \rightarrow \eta l_B$ ($\eta$ is a constant) simultaneously, the scaling symmetry recovers. Thus, the magnetic dependence of the spectrum exhibit the characteristics of the Efimov-like quasibound states. This can be regarded as a magnetic analogy of the Efimovian expansion\cite{Efimov_Dyn1}.

\begin{figure}[h] 
    \includegraphics[width=8.0cm,height=7.9cm]{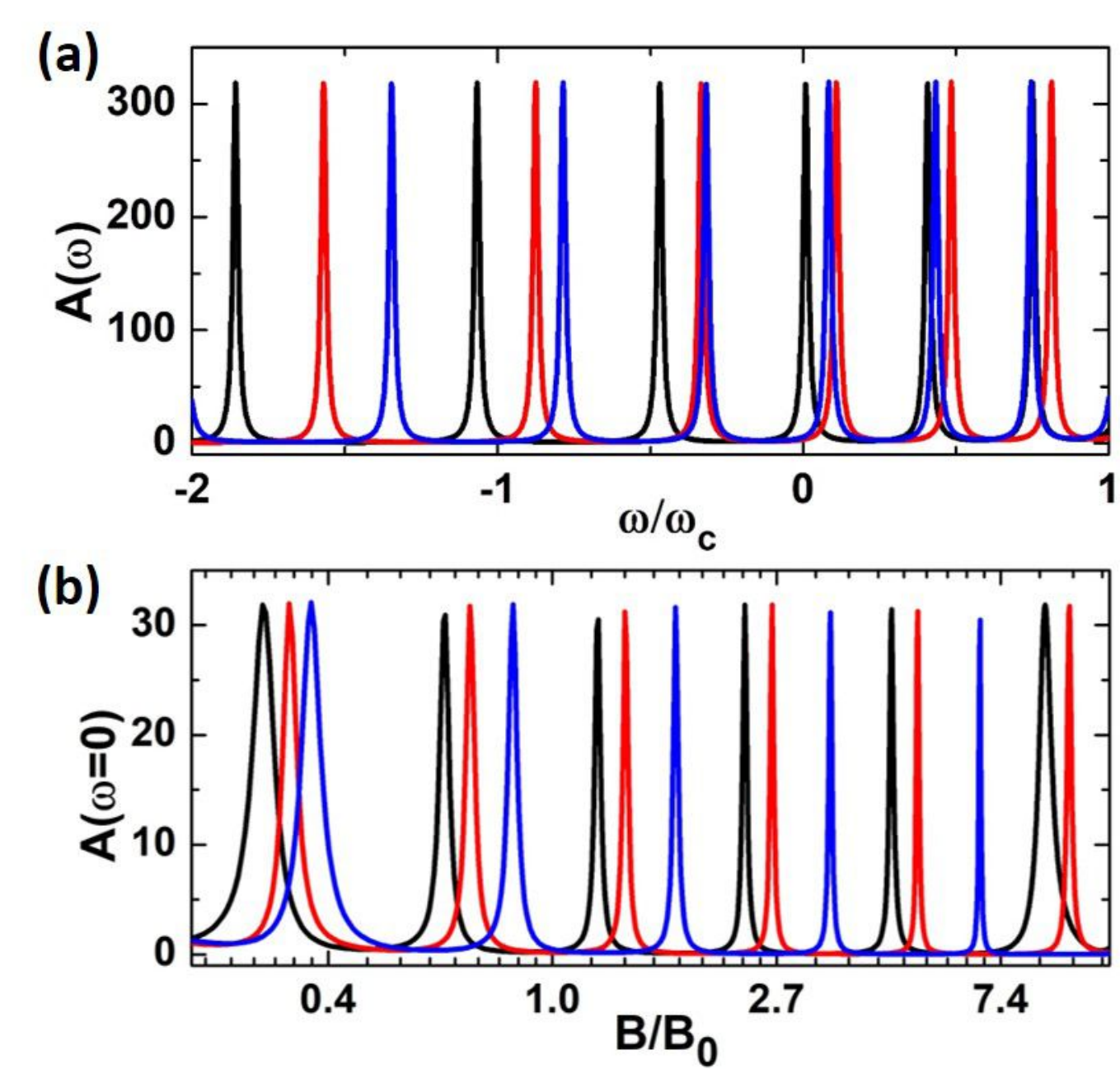}
    \caption{(Color Online). Spectra of the 2D massless Dirac electrons at different channals $m=0$ (black),$1$ (red) and $4$ (blue). The same parameters are used as Fig.~\ref{results2Dsf} and \ref{results2Dsfb}. (a) the spectrum of three different channels showing that the binding energies vary with the channels. (b) Magnetic dependence of the density of states $A(\omega)$ at the Dirac point. They exhibit different scaling factors for three channels.}
     \label{results2Dsfm}
\end{figure}

For different channels, the scaling factors are different. Furthermore, the energy locations of quasibound states are also different, even with the same short-range boundary condition (see Fig.~\ref{results2Dsfm}). Thus, in zero magnetic field, the quasibound states do not obey a discrete geometric scaling law when more than one channels are occupied, although they do for each channel respectively. The corresponding magnetic oscillation of the spectrum does not exhibit a DSS.

\section{Discrete scaling symmetry in 3D Dirac semimetals}

Similarly, in 3D, if we assume $\psi(\mathbf{r})=(u_1(r)\mathcal{Y}_{j-1/2}^{jm}(\theta,\varphi),-i u_2(r)\mathcal{Y}_{j+1/2}^{jm}(\theta,\varphi) )^T$, the radial equation can be expressed as\cite{Sakurai}
\begin{equation}
\frac{d}{dr}
  \begin{pmatrix}
   u_1(r)  \\
   u_2(r) 
 \end{pmatrix}=
 \begin{pmatrix}
   \frac{\lambda-1}{r} & -(\frac{E}{\hbar v_F}+\frac{Z\alpha}{r}) \\
   \frac{E}{\hbar v_F}+\frac{Z\alpha}{r}  &  -\frac{\lambda+1}{r}
 \end{pmatrix}
 \begin{pmatrix}
   u_1(r)  \\
   u_2(r) 
\end{pmatrix}
\end{equation} 
Here, $\lambda=\pm(j+1/2)$ is an integer and $\mathcal{Y}_{j\mp 1/2}^{jm} (\theta,\varphi)$ is the spinor function defined as
\begin{equation}
 \mathcal{Y}_{l=j\mp 1/2}^{jm}=\frac{1}{\sqrt{2l+1}}
 \begin{pmatrix}
   \pm\sqrt{l\pm m+\frac{1}{2}}Y_l^{m-1/2}(\theta,\varphi)  \\
   \sqrt{l\mp m+\frac{1}{2}}Y_l^{m+1/2}(\theta,\varphi) 
 \end{pmatrix}
\end{equation}
with $Y_l^{m\mp 1/2} (\theta,\varphi)$ being the spherical harmonics.

For $|Z\alpha| > |\lambda|$, there also exist a series of Efimov-like quasibound states in each channel $\lambda$, with a scaling factor $e^{\pi/\sqrt{(Z\alpha)^2-\lambda^2}}$\cite{Wang, DiracEfimov3, DiracEfimov4}. When a magnetic field (along $z$-axis) is applied, the rotation symmetry is broken and the total angular momentum $j$ is not conserved ($m$ remains a good quantum number due to the remaining rotation symmetry in $xy$-plane). Thus different channels couple with each other. Since the scaling factors vary for different channels, all quasi-bound states do not possess a DSS. The magnetic oscillation of the spectrum also breaks the original DSS. However, in the quantum limit (as in the experiment\cite{Wang}), the 3D system becomes quasi-1D (parallel to the direction of the magnetic field). The total angular momentum $j$ is determined for shallow quasibound states by the Landau level, where decoupled channels can be denoted by $m$ as in 2D. For the direction along $z$-axis, the magnetic field just supplies a magnetic length scale in the transversal plane, which makes the system like a tube. In a semi-classical way, the Coulomb potential becomes $V(\mathbf{r})=-Z\alpha/\sqrt{l_B^2+z^2}$ and it does not affect the 1D massless Dirac fermions significantly, similar to Klein tunneling. In the $xy$-plane, when one Dirac electron is close to the impurity, it can resonantly scatter with one nearby quasibound state of the impurity just as in 2D, and the energy can be modified drastically. Since the quasibound states can be shifted by the magnetic field, the spectrum will oscillate in a log-periodic way, which qualitatively agrees well with the observed magnetoresistance oscillation\cite{Wang}.

\section{Discussion and Outlook}
In this paper, we show that a discrete scaling symmetry can be displayed in both 2D and 3D Dirac semimetals, with attractive charged impurities in a magnetic field. Although they both result from the Efimov-like quasibound states in zero magnetic field, their detailed conditions are quite different. In 2D, as the magnetic field is applied, the magnetic oscillation of the spectrum exhibits a DSS, by the interplay between quasibound states and Landau levels.  However, in 3D, before the system enters the quantum limit, different channels couple with each other and the magnetic oscillation do not show a DSS due to the lack of a unique scaling factor. Nevertheless, in the quantum limit, the 3D system becomes quasi-1D and the channels are decoupled for shallow quasibound states. The magnetic dependence of the spectrum around impurities can exhibit a DSS.

In order to observe this phenomenon in experiments, there are several key factors. Firstly, there should be attractive charged impurities (or other ways to offer an attractive Coulomb potential) in Dirac semimetals, either electrons with positive charged impurities or holes with negative charged impurities. Furthermore, the electric charge of the impurity should be bigger than a critical value, so as to possess a scaling factor $e^{\pi/\sqrt{(Z\alpha)^2-1/4}}$ ($m=0$) for 2D and $e^{\pi/\sqrt{(Z\alpha)^2-1}}$ ($\lambda=1$) for 3D. Secondly, the magnetic field should be appropriate to shift the quasibound states significantly. In particular, for 3D, it should be strong enough to drive the system into quasi-1D, so as to decouple the original channels. Thirdly, compared with the Landau level's orbit, the density of current carriers (electrons or holes) should be low, to make sure that only one channel is occupied. Otherwise, other channels' contribution can elude the observation. The conditions of the recent experiment\cite{Wang} are consistent with the above standards and our results are in good agreement with their observations qualitatively. In order to compare quantitatively, another length scale should be introduced in the direction parallel to the magnetic field, for example, the distance between neighbouring impurities, to estimate how much the density of states (or the conductance) can be modified. It will involve the interference between impurities. Furthermore, the electron-electron interaction could play an important role in real systems and it will be interesting to see how it will change the quasibound states and the magnetic dependence of the spectrum. These works can be studied in the future.

\begin{acknowledgements} 
 We thank Haizhou Lu, Haiwen Liu, Pengfei Zhang, Xin Chen, Zhigang Wu and Hui Zhai for helpful discussions. This work was inspired by Haiwen Liu's talk at IAS, Tsinghua University.
\end{acknowledgements}

\end{document}